\documentclass[12pt]{iopart}

\usepackage{iopams}
\usepackage{epsfig}
\usepackage{color}

\newcommand{\BEQ}{\begin{equation}}     
\newcommand{\BEA}{\begin{eqnarray}}
\newcommand{\EEQ}{\end{equation}}       
\newcommand{\EEA}{\end{eqnarray}}
\newcommand{\D}{{\rm d}}                
\newcommand{\II}{{\rm i}}               

\newcommand{\appsektion}[1]{\setcounter{equation}{0} \section*{Appendix: #1}
\renewcommand{\theequation}{A\arabic{equation}}
              \renewcommand{\thesection}{A} }

\begin{document}
\title[Kinetics of a non-glauberian Ising model]{
Kinetics of a non-glauberian Ising model: global observables and exact results}
\author{Sreedhar B. Dutta$^a$, Malte Henkel$^{b}$ and Hyunggyu Park$^a$}
\address{$^a$School of Physics, Korea Institute for Advanced Study,\\ 
87 Hoegiro, Dongdaemun-gu, Seoul 130-722, South Korea\\
$^b$Groupe de Physique Statistique, D\'epartement 1: Physique de la Mati\`ere et des Mat\'eriaux, Institut Jean Lamour,\footnote{Laboratoire associ\'e au CNRS UMR7198}
CNRS - Nancy Universit\'e - UPV Metz, \\
B.P. 70239, F -- 54506 Vand{\oe}uvre l\`es Nancy Cedex, France
}
\begin{abstract}
We analyse the spin-flip dynamics in kinetic Ising chains with Kimball-Deker-Haake (KDH) transition rates, and evaluate exactly the evolution of global quantities like magnetisation and its fluctuations, and the two-time susceptibilities and correlations of the global spin and the global three-spin. Information on the ageing behaviour after a quench to zero temperature is extracted. 
\end{abstract}

\section{Introduction}

Kinetic Ising models have been studied for nearly half a century, and have helped in addressing various fundamental issues in non-equilibrium physics. In spite of their simplicity, they have led to our understanding in various aspects, ranging from the relaxation mechanisms for equilibration to the collective behaviour in 
non-equilibrium systems. Unfortunately, even these simple models could only be solved exactly in very few cases, and one such case is the celebrated Glauber-Ising model~\cite{Glauber63}. 

The Glauber-Ising model is driven by spin-flip transitions and shows critical slowing-down near zero temperature with a dynamic exponent $z=2$ in one dimension. Variational techniques and renormalisation-group methods suggest that spin-flip dynamics leads to $z=2$, provided the energy-conserving flips are not excluded at any temperature~\cite{Deker79,Achiam80}. Kimball~\cite{Kimball79}, and Deker and Haake~\cite{Deker79} (KDH) studied a different kinetic Ising model, also driven by spin-flip transitions, but with rates which differ from those of the Glauber-Ising model, and find that $z=4$ near zero temperature. Haake and Thol~\cite{Haake80} further found a continuous set of single-spin-flip Ising models near zero temperature with $z$ ranging from 2 to 4.\footnote{Their exact results are easily reproduced by heuristic arguments about the movement of domain walls \cite{Cordery81} and have been extended to $q$-state Potts models \cite{Droz86}.} These models with $z \ne 2$ received a considerable amount of attention in the past, see e.g. \cite{Cornell} for reviews. 

The equations of motion for total spin and three-spin in the KDH-Ising model are known to form a closed set of linear equations, and hence the global magnetisation could be evaluated exactly \cite{Kimball79,Deker79}. Since solvability for KDH dynamics is quite restricted, it has not been explored to the same extent as the Glauber-Ising model. Recently, two-time correlators and responses have been evaluated in the Glauber-Ising model to study its ageing properties \cite{Godreche00a, Lippiello00, Henkel04} after a quench to temperature $T=0$. It has been shown that the two-time space-time-dependent spin-spin correlator, $C(t,s;r,r')=\langle\sigma_r(t)\sigma_{r'}(s)\rangle$, and the (linear) response function, $R(t,s;r,r')=\left.\frac{\delta \langle\sigma_r(t)\rangle}{\delta h_{r'}(s)}\right|_{h=0}$, satisfy the scaling forms, if $s$ {\em and} $t-s>0$ are sufficiently large (see e.g. \cite{Calabrese05} for a review)
\BEQ \label{gl:scal}
\hspace{-2truecm}C(t,s;r,r') = s^{-b} F_C\left(\frac{r-r'}{(t-s)^{1/z}}, \frac{t}{s}\right) 
\;\; , \;\;
R(t,s;r,r') = s^{-a-1} F_R\left(\frac{r-r'}{(t-s)^{1/z}}, \frac{t}{s}\right),
\EEQ
where $a,b$ are ageing exponents, $z$ is the dynamical exponent. Here spatial translation-invariance has been admitted. For $y=t/s\gg 1$, one  further expects $F_{C,R}(0,y)\sim y^{-\lambda_{C,R}/z}$, where $\lambda_{C,R}$ are the autocorrelation and autoresponse exponent, respectively. On the other hand, for correlations and
responses with respect to the initial state one expects, for sufficiently large times $t$ (see e.g. \cite{Calabrese05})
\BEQ \label{gl:scalinf}
\hspace{-2truecm}C(t,0;r,r') = t^{-\lambda_C/z} \Phi_C\left(\frac{r-r'}{t^{1/z}}\right) 
\;\; , \;\;
R(t,0;r,r') = t^{-\lambda_R/z} \Phi_R\left(\frac{r-r'}{t^{1/z}}\right)
\EEQ
Can one find similar non-equilibrium scaling forms in the KDH-Ising model~? And if so, what are the values of the exponents~?

The layout of the paper is as follows. In the next section we define the model and derive the equations of motion for one- and two-point functions for spin and three-spin variables. In section~\ref{global}, we derive exactly the susceptibilities and fluctuations of the global magnetisation and the global three-spin, and also discuss a dual description of the model as a reaction-diffusion process. We conclude in section~\ref{conclude}. In an appendix, details of the derivation of the equation of motion for global two-point correlators are presented. 

\section{Kinetic Ising chains}
\label{KIC}
In this section, we define kinetic Ising models which evolve under single spin-flip dynamics, and explicitly find the equations of motion for expectation values and correlation functions for spin and three-spin observables. 

\subsection{KDH dynamics}
\label{KHD}

We consider a kinetic Ising model on a chain $\Lambda$ with energy 
\begin{equation}
\label{energy}
{\cal H}[\sigma]=-J\sum_{n\in\Lambda} \sigma_n \sigma_{n+1}- \sum_n h_n \sigma_n, ~~ J > 0,
\end{equation}
where a spin configuration is denoted by $\sigma := \{\cdots, \sigma_{n-1}, \sigma_n, \sigma_{n+1}. \cdots\}$, and $\sigma_n=\pm 1$ is the Ising spin variable at lattice site $n$. The chain $\Lambda\subset\mathbb{Z}$ has $L$ sites and we shall take the thermodynamic limit $L \rightarrow \infty$ throughout. The system is assumed to be in contact with a heat-bath that induces only local spin-flip transitions, in other words, only single spins are flipped such that the transition rates depend just on the flipped spin $\sigma_n$ and its nearest neighbours $\sigma_{n\pm 1}$. A configuration $\sigma$, where the sign of the spin at site $n$ is flipped is denoted 
by $F_n\sigma := \{\cdots, \sigma_{n-1}, -\sigma_n, \sigma_{n+1}. \cdots\}$. We are interested in the following transition rates, whose form depends on the parameters
$\gamma,\delta$ 
\begin{eqnarray}
\label{Ws}
\lefteqn{ W(F_n\sigma|\sigma) = \alpha \left( 1 - \frac{\gamma}{2} \sigma_n\left(\sigma_{n-1} + \sigma_{n+1}\right) + 
\delta \sigma_{n-1}\sigma_{n+1}\right) } \nonumber \\
&-& \alpha \tanh(\beta h_n) \left(\sigma_n - \frac{\gamma}{2}\left(\sigma_{n-1} + \sigma_{n+1}\right) + 
\delta \sigma_{n-1} \sigma_n\sigma_{n+1}\right),
\end{eqnarray}
where $\beta$ is the inverse temperature and $\alpha$ is a normalisation constant. For vanishing fields $h_n=0$, these are the most general local transition rates  $\sigma \mapsto F_n\sigma$ which are left-right (parity)-symmetric and also invariant under reversal under all spins~\cite{Glauber63}. The condition of detailed balance implies that 
\begin{equation}
\label{gam} 
\gamma = (1+\delta)\tanh(2\beta J).
\end{equation} 
The field-dependent terms are the most simple ones which are also compatible with detailed balance (for example, they are used in \cite{Glauber63,Henkel04}, while in \cite{Godreche00a} a slight variant is studied). In what follows, we shall essentially restrict ourselves to the case $\gamma=  2 \delta$ which we shall call {\em KDH dynamics}. Explicitly, when combined with (\ref{gam}), the value of 
\BEQ
\delta = \frac{\gamma}{2} = \frac{\tanh 2\beta J}{2-\tanh 2\beta J} ~, 
\EEQ
and hence $\delta\to 1$ when the temperature $T=\beta^{-1}\to 0$. 
As we shall see, this choice produces a closed set of dynamical equations for some global observables. This observation goes back to Kimball \cite{Kimball79}, and Deker and Haake \cite{Deker79}. The case with $\delta=0$ in equation (\ref{Ws}) is the usual Glauber dynamics \cite{Glauber63}.

The master equation with only single spin-flip transitions is given by
\begin{equation}
\label{master}
\frac{\partial}{\partial t} P(\sigma,t) = -\sum_n \left[ W(F_n\sigma|\sigma)P(\sigma,t) 
-  W(\sigma|F_n\sigma)P(F_n\sigma,t) \right],
\end{equation}
and, following Glauber \cite{Glauber63}, can be rewritten as
\begin{equation}
\label{master-2}
\frac{\partial}{\partial t} P(\sigma,t) = -\sum_n \sigma_n \sum_{\sigma'_n=\pm1} \sigma'_n
\left[ W(F_n\sigma|\sigma)P(\sigma,t) \right]_{\sigma_n \rightarrow \sigma'_n}.
\end{equation}
Hence the dynamical equation for the $N$-point function is given by
\begin{equation}
\label{N-pt}
\frac{\partial}{\partial t} \langle \sigma_{n_1} \cdots \sigma_{n_N}\rangle_t
 = -2 \left\langle \sigma_{n_1} \cdots \sigma_{n_N}
\sum_{i=1}^N W(F_{n_i}\sigma|\sigma) \right\rangle_t ,
\end{equation}
where $n_1, \cdots, n_N$ are $N$ non-coinciding sites, and $\langle \cdots \rangle_t := \sum_{\sigma} \cdots P(\sigma,t)$.

Similarly, we can obtain equations of motion for two-time quantities from the conditional probability measure 
$P(\sigma,t| \widetilde{\sigma},s)$ for a configuration to be $\sigma$ at time $t$ conditioned to the configuration $\widetilde{\sigma}$ at an earlier time $s$.  The two-time correlation function of  $X :=  \sigma_{n_1} \cdots \sigma_{n_N}$ and $Y:=\sigma_{n_1} \cdots \sigma_{n_M}$ for $t > s$ is defined as
\begin{equation}
\label{2-time-fn}
\langle X(t)Y(s) \rangle := \sum_{\sigma,\widetilde{\sigma}} X \widetilde{Y}P(\sigma,t| \widetilde{\sigma},s)
\end{equation}
where $\widetilde{Y} := \widetilde{\sigma}_{n_1} \cdots \widetilde{\sigma}_{n_M}$.
The dynamic equation for $P(\sigma,t| \widetilde{\sigma},s)$ is similar to equation~(\ref{master-2}) with $P(\sigma,t)$ replaced by $P(\sigma,t| \widetilde{\sigma},s)$, and hence the equation of motion for two-time correlation functions is similar to equation~(\ref{N-pt}).

\subsection{Equations of motion}
\label{e.o.m}
Using equations (\ref{Ws}) and (\ref{N-pt}), and re-scaling time such that $\alpha=1/2$, we get the following equation of motion for the expectation value of the spin: 
\begin{eqnarray}
\label{1-pt}
\frac{\partial}{\partial t} \langle \sigma_n\rangle
&=& -\langle \sigma_n \rangle + \frac{\gamma}{2} \langle \sigma_{n-1} + \sigma_{n+1} \rangle
- \delta \langle q_n \rangle  
\nonumber \\
& & +\beta h_n \left( 
1 -\frac{\gamma}{2} \left\langle \sigma_{n}\left(\sigma_{n-1}+\sigma_{n+1} \right) \right\rangle 
+ \delta \langle  \sigma_n q_n \rangle \right),
\end{eqnarray}
to linear order in the field $h_n$, where the three-spin variable $q_n$ is defined as
\begin{equation}
\label{q-def}
q_n := \sigma_{n-1}\sigma_n\sigma_{n+1}
\end{equation}
Similarly, the equation of motion for the average three-spin 
$\langle q_n \rangle$ is 
\begin{eqnarray}
\label{3-pt}
\hspace{-1truecm}\frac{\partial}{\partial t} \langle q_n\rangle
&=& -3\langle q_n\rangle + \gamma \langle \sigma_{n-1} + \sigma_{n+1} \rangle - \delta \langle \sigma_n \rangle  
+ \delta \left\langle  A_n \right\rangle
\nonumber \\
& & +\beta\! \sum_{m=0,\pm 1} \! h_{n+m} \left\langle q_n\left(
\sigma_{n+m} - \frac{\gamma}{2} (\sigma_{n+m-1} +  \sigma_{n+m+1}) + \delta q_{n+m} \right) \right\rangle ,
\end{eqnarray}
again to linear order in $h_n$, and where
\begin{equation}
\label{An}
\hspace{-1.5truecm}A_n := \left[ \frac{\gamma}{2\delta} \sigma_{n-2} \sigma_n\sigma_{n+1} -\sigma_{n-1} \sigma_{n+1}\sigma_{n+2} \right] 
+ \left[ \frac{\gamma}{2\delta} \sigma_{n-1} \sigma_n\sigma_{n+2}
 - \sigma_{n-2} \sigma_{n-1}\sigma_{n+1}\right]  .
\end{equation}
In the absence of magnetic field $h_n$, and under the assumption of translation invariance of the 3-point functions, $\langle A_n \rangle=0$ if $\gamma=2\delta$. Then (\ref{1-pt}) and (\ref{3-pt}) form a closed set of linear equations~\cite{Kimball79,Deker79}.

We now write down the equations for equal-time correlation functions when $h_n=0$. The evolution of spin-spin correlation function $\langle \sigma_n \sigma_m\rangle$ for $n \ne m$, obtained from equations (\ref{Ws}) and 
(\ref{N-pt}), is governed by
\begin{eqnarray}
\label{s-s}
\frac{\partial}{\partial t} \langle \sigma_m \sigma_n \rangle
= &-&2 \langle \sigma_m \sigma_n \rangle 
+ \frac{\gamma}{2} \langle \sigma_m( \sigma_{n-1} + \sigma_{n+1}) + \sigma_n( \sigma_{m-1} + \sigma_{m+1}) \rangle 
\nonumber \\
 &-&\delta \langle \sigma_m q_n + \sigma_n q_m\rangle . 
\end{eqnarray}
The correlation function $\langle q_n \sigma_m\rangle$, for $|m-n| > 1 $, satisfies the dynamic equation,
\begin{eqnarray}
\label{q-s}
\frac{\partial}{\partial t} \langle \sigma_m q_n\rangle
= &-& 4 \langle \sigma_m q_n\rangle 
+ \frac{\gamma}{2} \langle 2\sigma_m( \sigma_{n-1} + \sigma_{n+1}) + q_n( \sigma_{m-1} + \sigma_{m+1}) \rangle 
\nonumber \\
 &-&\delta \langle \sigma_m \sigma_n + q_m q_n \rangle  
 + \delta \langle \sigma_m A_n \rangle ,
\end{eqnarray}
where $A_n$ is given in equation (\ref{An}). The dynamics of the auto-correlations 
$ \langle \sigma_{n} q_n\rangle = \langle \sigma_{n-1} \sigma_{n+1} \rangle$ 
and that of the nearest-neighbour correlations 
$\langle \sigma_{n \pm 1} q_n\rangle = \langle \sigma_n \sigma_{n\mp 1} \rangle$ 
are determined from equation (\ref{s-s}). We further obtain the equation of motion for the correlation function 
$\langle q_n q_m\rangle$, for $|n-m|> 2$, to be
\begin{eqnarray}
\label{q-q}
\frac{\partial}{\partial t} \langle q_m q_n\rangle
= &-&6 \langle q_m q_n \rangle 
+ \gamma 
\langle q_m(\sigma_{n-1}+\sigma_{n+1})+q_n( \sigma_{m-1}+\sigma_{m+1}) \rangle  
\nonumber \\
&-&\delta \langle q_m \sigma_n + q_n \sigma_m \rangle + \delta \langle q_m A_n 
+ q_n A_m \rangle . 
\end{eqnarray}
The dynamics of the nearest-neighbour correlations 
$ \langle q_n q_{n+1}\rangle = \langle \sigma_{n-1} \sigma_{n+2} \rangle$ is 
given by equation (\ref{s-s}), while that of the next-nearest one, 
$\langle q_{n-1} q_{n+1}\rangle = \langle\sigma_{n-2}\sigma_{n-1}\sigma_{n+1} \sigma_{n+2}\rangle$, 
can be obtained from equations (\ref{Ws}) and (\ref{N-pt}), and is given by
\begin{eqnarray}
\label{q-q-2}
\frac{\partial}{\partial t} \langle q_{n\!-\!1} q_{n\!+\!1}\rangle
= &-&4 \langle q_{n-\!1} q_{n+1}\rangle 
+ \gamma 
\langle \sigma_{n\!+\!1}\sigma_{n\!+\!2}\!+\!\sigma_{n\!-\!2}\sigma_{n\!-\!1} \rangle  
+ 2\delta  \langle \sigma_{n\!-\!2} \sigma_{n\!+\!2}\rangle 
\nonumber \\
&-&\delta \langle \sigma_{n\!-\!1} q_{n\!+\!1}\! +\! q_{n\!-\!1} \sigma_{n\!+\!1} \rangle 
+ \delta \langle q_{n\!-\!1} A_{n\!+\!1} \!+ \!q_{n\!+\!1} A_{n\!-\!1} \rangle . 
\end{eqnarray}
Since $A_n$ appears in the equations (\ref{3-pt},\ref{s-s},\ref{q-s},\ref{q-q},\ref{q-q-2}), 
these equations are in general not closed and are therefore unsolvable.  

\section{Global observables}
\label{global}
We now derive exact results for the behaviour of {\em global} magnetisation,  response and correlation functions under KDH dynamics, when $\gamma = 2 \delta$ is chosen.  

\subsection{Magnetisation}
\begin{figure*}[t]
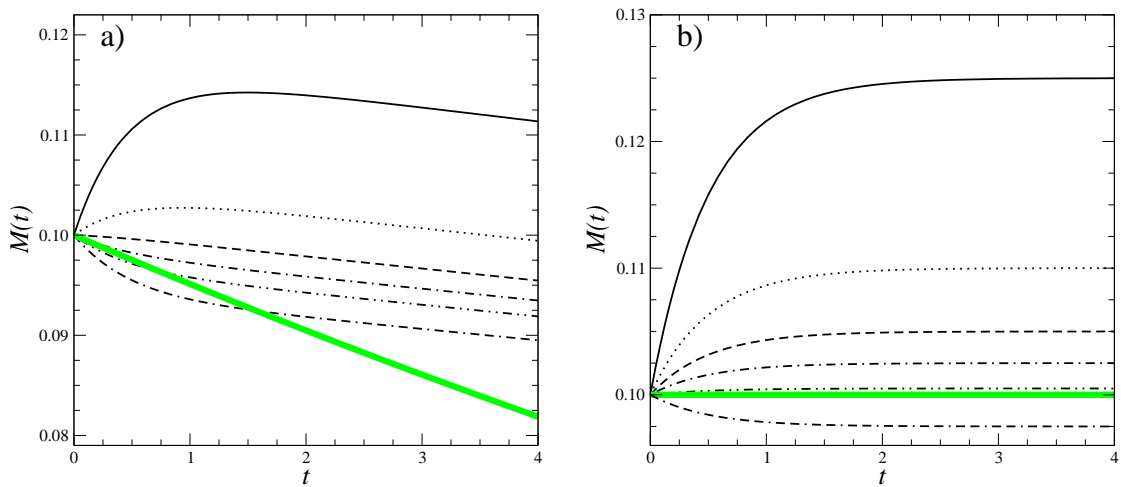

\begin{center}
     \includegraphics[width=2.8in,angle=0,clip=]{mag_KDHglau_h.eps} ~~
     \includegraphics[width=2.8in,angle=0,clip=]{mag_KDHglau_c.eps}
\end{center}
\caption[Fig 1]{Time-dependence of the global magnetisation $M(t)$ in the $1D$ Ising model with KDH dynamics for the initial values $T(0)/M(0)=[0.50,0.80,0.90,0.95,0.99,1.05]$ from top to bottom and for (a) $\delta=0.90476\ldots$ and (b) $\delta=1$. The thick grey lines give the time-dependent global magnetisation of the Glauber-Ising model with the same
values of $\beta J$. 
\label{Fig1}
}
\end{figure*}

The global average of spin, $M(t)$, and that of three-spin, $T(t)$, are defined as
\BEQ
\label{M}
M(t) := \frac{1}{L}\sum_{n} \langle \sigma_n \rangle_t \;\; , \;\; 
T(t) := \frac{1}{L}\sum_{n} \langle q_n \rangle_t ,
\EEQ
where $L$ is the number of lattice sites. The equations of motion for these observables follow from 
equations (\ref{1-pt},\ref{3-pt}) and read for $h_n=0$ 
\begin{eqnarray}
\label{M-T-dyn}
\frac{\D}{\D t} 
\left( \begin{array}{c} 
M(t)\\T(t)
\end{array} \right)
= \left( \begin{array}{cc} 
2\delta-1& - \delta \\
3\delta &-3
\end{array} \right)
\left( \begin{array}{c} 
M(t)\\T(t)
\end{array} \right) .
\end{eqnarray}
Hence the global magnetisation $M(t)$ and the global three-spin average $T(t)$ are 
\BEA
\label{M-t}
M(t) &=& \frac{1}{2\Delta} \left( \left( \alpha_{+}M(0) - \delta T(0) \right) e^{-\lambda_{-}t} -  
\left( \alpha_{-}M(0) - \delta T(0) \right)e^{-\lambda_{+}t} \right) ,
\\
T(t) &=& \frac{1}{2\Delta\delta} \left( \alpha_{-}\left( \alpha_{+}M(0) - \delta T(0) \right) e^{-\lambda_{-}t} -  
\alpha_{+}\left( \alpha_{-}M(0) - \delta T(0) \right)e^{-\lambda_{+}t} \right) ,
\nonumber
\EEA
where 
\BEQ
\label{eigen-lam}
\lambda_{\pm}= 2-\delta \pm \Delta \;\; , \;\; \Delta=(1+2\delta-2\delta^2)^{1/2} 
\;\; , \;\; 
\label{alph}
\alpha_{\pm} = 1 + \delta \pm \Delta .
\EEQ
In contrast to Glauber dynamics~\cite{Glauber63}, the magnetisation can exhibit a 
non-monotonous behaviour due to the presence of two time-scales, see 
figure~\ref{Fig1}. It has a monotonous decay when the initial conditions are such 
that $\alpha_{-}\delta^{-1} \le  T(0)/M(0) \le \alpha_{+}\delta^{-1} $. Otherwise, 
the global magnetisation first increases or decreases rapidly until a time $t^*$ and 
then decays more slowly with the finite time constant $\lambda_{-}$. 
Thus there is an initial increase in correlations with time before they begin to 
decay. We also see from figure~\ref{Fig1}b that in the zero-temperature limit ($\delta\to 1$) the stationary value $M(\infty)$ no longer equals $M(0)$ in general, in contrast to the case of Glauber dynamics. The crossover time is given by
\begin{equation}
\label{t*}
t^* = \frac{1}{2\Delta} \ln\left[\frac{(\alpha_{-} M(0)- \delta T(0))\lambda_{+}}
{ (\alpha_{+} M(0)- \delta T(0))\lambda_{-}} \right],
\end{equation}
and depends on the ratio $M(0)/T(0)$ and the temperature. This time-scale diverges either with fine-tuned initial conditions, $ \alpha_{+}M(0) - \delta T(0) \approx 0$, or by approaching low temperatures where $\delta \approx 1$.

Now the equilibrium correlation-length $\xi$ at low temperatures behaves as
\begin{equation}
\label{corr-len}
\xi^{-1} = -\ln\tanh(\beta J) \approx 2e^{-2\beta J} +\frac{2}{3} e^{-6\beta J}.
\end{equation}
Using the equation (\ref{corr-len}) and the relation (\ref{gam}) for $\gamma=2\delta$, we get
\begin{equation}
\label{del-lw}
\delta \approx 1 -\xi^{-2} + \frac{11}{12} \xi^{-4}.
\end{equation}
In this limit, for generic initial conditions which give rise to non-monotonous behaviour, the time-scale $t^*\approx 2\ln\xi$ diverges logarithmically. Hence at low temperatures, the correlations gradually increase for a long transient time $t^*$. After the crossover, for $t \gg t^*$ the relevant time-scale is the 
relaxation time associated with $\lambda_{-}$ and is given by
\begin{equation}
\label{t-}
\tau_{-} = \lambda_{-}^{-1} \approx \frac{2}{3} \xi^{4},
\end{equation}
and the dynamical exponent is~\cite{Deker79} 
\begin{equation}
\label{z}
z=4.
\end{equation}
The reason for this exceptional value of the dynamical exponent is more transparent in a dual description, see below.

Finally, we observe that if we choose $\gamma=-2\delta$, then the staggered quantities 
\begin{equation} 
M_s(t) := \frac{1}{L} \sum_{n} \langle (-1)^n \sigma_n \rangle_t  \;\; , \;\;
T_s(t) := \frac{1}{L} \sum_{n} (-1)^n \langle q_n \rangle_t 
\end{equation}
satisfy the {\em same} system (\ref{M-T-dyn}) of equations of motion as those followed by the pair
$(M(t),T(t))$ in the case of KDH dynamics with $\gamma=+2\delta$. Therefore, the conclusions reached for the ferromagnetic KDH model can be carried over to its anti-ferromagnetic analogue.

\subsection{Dual description: reaction-diffusion processes}
The nature of the relaxation dynamics along KDH line near zero temperature is more revealing in the particle-picture \cite{Racz85} of the Ising chain. In this dual description a particle ($A$) is associated 
with a kink (either $\uparrow \downarrow $ or $\downarrow \uparrow $) and a vacancy ($\emptyset$) is associated with its absence (either $\uparrow \uparrow $ or $\downarrow  \downarrow$). Thus the spin-flip 
transitions are identified with the reaction-diffusion processes of the particles as shown in table~\ref{tab1}. A global reversal of the spins leaves the corresponding particle configuration unchanged, and hence 
only half of the spin-flip processes are listed in table~\ref{tab1}. 

\begin{table}
\begin{center}\begin{tabular}{|ccc|} \hline
process & rate & dual process \\ \hline 
$\uparrow \downarrow \uparrow \longrightarrow \uparrow \uparrow \uparrow $ & $\alpha(1+\gamma+\delta)$ & $AA \longrightarrow \emptyset\emptyset$ \\[1.5mm]
$\uparrow \uparrow \uparrow \longrightarrow \uparrow \downarrow \uparrow$ & $\alpha(1-\gamma+\delta)$  & $\emptyset\emptyset \longrightarrow AA$ \\[1.5mm]
$ \uparrow \uparrow \downarrow \longrightarrow \uparrow \downarrow\downarrow$ & $\alpha(1-\delta)$  & $\emptyset A \longrightarrow A \emptyset$ \\[1.5mm]
$\uparrow \downarrow\downarrow \longrightarrow \uparrow\uparrow\downarrow$ & $\alpha(1-\delta)$  & $A\emptyset \longrightarrow \emptyset A$ \\ \hline
\end{tabular}
\caption[tab1]{Elementary spin-flip processes  and their rates along with the dual description. 
\label{tab1}
}
\end{center}
\end{table}

\begin{figure}[t]
\begin{center}
\includegraphics[width=6cm]{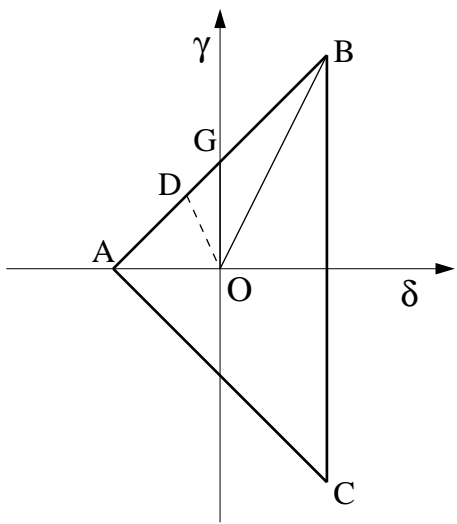}
\caption{\label{Fig2} The parameter space $(\delta,\gamma)$ of spin-flip 
transitions is confined to the domain $ABC$; $AB$ is $T=0$ line, $OB$ is 
KDH line, $OG$ is Glauber line, $BC$ is $\delta=1$, and $OD$ is 
$\gamma=-2\delta$.}
\end{center}
\end{figure}

Non-negative transition rates are found for the region ABC in the parameter space $(\delta, \gamma)$ as shown 
in figure~\ref{Fig2}. The line AB is $\gamma=1+\delta$ and corresponds to zero temperature, while the $\delta$-axis corresponds to infinite temperature. The line OB is the KDH line $\gamma=2\delta$, and as we move from O to B, the temperature drops from infinity to zero. The line BC is $\delta=1$ where the diffusion process 
$A\emptyset \longleftrightarrow \emptyset A$ is completely suppressed. Since the rate of diffusion $D =  \alpha(1-\delta)$, the particles or the kinks diffuse a distance $\xi$ in time $\tau$, where 
\begin{equation}
\label{diff}
\tau = \frac{\xi^2}{D} = \frac{\xi^2}{\alpha(1-\delta)}.
\end{equation}
As we approach the zero temperature the detailed-balance condition implies, as can be seen from equations (\ref{gam}) and (\ref{corr-len}), that $\gamma \approx (1+\delta)(1-\xi^{-2}/2)$. The parameter $\delta$ independently depends on temperature, and along KDH line near zero temperature $\delta \approx 1-\xi^{-2}$. Now near the point $(\delta,\gamma)=(1,2)$ the most dominant process is the pair-annihilation $AA \longrightarrow \emptyset\emptyset$, and for times larger than the corresponding transient time-scale the system reaches 
one of the many absorbing states of isolated particles. The remaining slow processes can induce diffusion and, from equation (\ref{diff}), we get the typical time needed to relax towards a state with domains of typical size $\xi$ as $\tau \approx \xi^4$~\cite{Cordery81, Haake80}.

The nature of these dynamical classes can be qualitatively understood by 
comparing the transition rates of energy-conserving flips 
$A\emptyset \longrightarrow \emptyset A$ with energy-costing flips 
$\emptyset\emptyset \longrightarrow AA$. For the KDH choice 
$\gamma=2\delta$, the rates for energy-conserving-flips and for energy-costing-flips 
are the same $\sim \xi^{-2}$, hence a large number of absorbing 
states are accessed, unlike the other near-equilibrium kinetic Ising 
classes with $z=2$. The number of absorbing states for large $L$ is of 
the order $((1+\sqrt{5})/2)^L$ \cite{Carlon01}, while there are only two 
zero-temperature equilibrium states.   

A similar analysis can be made in anti-ferromagnetic Ising chains for 
the choice $\gamma=-2\delta$. Along the line OD in figure~\ref{Fig2} 
the staggered magnetisation can be exactly evaluated. The relaxation 
time $\tau_s$ for such a configuration at low temperatures can be easily 
estimated in the dual picture, and in ferromagnetic models  $\tau_s \approx \xi^2$, 
while in anti-ferromagnetic models $\tau_s \approx \xi^4$.

\subsection{Global responses and correlations}
We now evaluate the susceptibilities and fluctuations of total spin and three-spin. 
The responses of the magnetisation and of total three-spin to a uniform 
time-dependent magnetic field $h(s)$ are defined as
\BEQ
\label{glob-R}
R(t,s) := \frac{1}{\beta L} \sum_n \left. \frac{\delta \langle\sigma_{n} \rangle_t}{\delta h(s)}  \right|_{h=0} \;\; , \;\;
\label{glob-Q}
Q(t,s) := \frac{1}{\beta L} \sum_n \left. \frac{\delta \langle q_{n} \rangle_t}{\delta h(s)}  \right|_{h=0} .
\EEQ
The fluctuations of the global spin and three-spin are described by
\BEQ
\label{glob-Cm}
C^{gf}_m(t) := \frac{1}{L} \sum_{n} C^{gf}_{n,n+m}(t) \;\; , \;\; 
\label{glob-C}
C^{gf}(t) := \frac{1}{L^2}\sum_{m,n} C^{gf}_{m,n}(t) ,
\EEQ
where $g$ and $f$ are either $\sigma$ or $q$, and the equal-time correlation functions are given by
\BEQ
\label{C-ss}
C^{\sigma \sigma}_{m,n}(t) :=  \langle \sigma_m \sigma_n \rangle_t \;\; , \;\;
\label{C-sq}
C^{q \sigma }_{m,n}(t) := \langle q_m \sigma_n \rangle_t \;\; , \;\; 
\label{C-qq}
C^{q q}_{m,n}(t) := \langle q_m q_n \rangle_t .
\EEQ

\subsubsection{Response functions:}
The equations of motion for $R(t,s)$ and $Q(t,s)$ are obtained from equations (\ref{1-pt}) and (\ref{glob-Q}). They read 
\begin{equation}
\label{glob-s-res}
\hspace{-0.2truecm}\frac{\partial}{\partial t} R(t,s) = (2\delta -1) R(t,s) -\delta Q(t,s) 
\end{equation}
for $t > s$, while for $t=s$ it is
\begin{equation}
\label{glob-s-res-in}
\hspace{-0.2truecm}R(s,s) = 1 - 2 \delta C^{\sigma \sigma}_{1}(s) + \delta C^{\sigma \sigma}_{2}(s).
\end{equation}
Similarly, we obtain the equation of motion for $Q(t,s)$ from equations (\ref{3-pt}) and (\ref{glob-Q}) and have for $t>s$ 
\begin{equation}
\label{glob-q-res}
\hspace{-0.2truecm}\frac{\partial}{\partial t} Q(t,s) = 3 \delta R(t,s) - 3 Q(t,s) ,
\end{equation}
and for $t=s$ the initial condition
\begin{equation}
\label{glob-q-res-in}
\hspace{-0.2truecm}Q(s,s) =  \delta + 2(1\!-\!\delta) C^{\sigma \sigma }_{1}(s) + (1\!-\!2\delta) 
C^{\sigma \sigma }_{2}(s) + 2\delta C^{\sigma\sigma}_{3}(s) -2\delta C^{q \sigma }_{2}(s) .
\end{equation}
The equations (\ref{glob-s-res}, \ref{glob-q-res}) are similar to (\ref{M-T-dyn}), 
and we find 
\begin{eqnarray}
\label{glob-R-soln}
\hspace{-0.2truecm}R(t,s)= \frac{1}{2\Delta} \left( A_{-}(s) e^{-\lambda_{-}(t-s)} - A_{+}(s) e^{-\lambda_{+}(t-s)} \right) ,\\
\label{glob-Q-soln}
\hspace{-0.2truecm}Q(t,s)= \frac{1}{2\Delta \delta} \left(\alpha_{-} A_{-}(s) e^{-\lambda_{-}(t-s)} 
- \alpha_{+}A_{+}(s) e^{-\lambda_{+}(t-s)} \right),
\end{eqnarray}
where $A_{\mp}(s) = \alpha_{\pm} R(s,s)- \delta  Q(s,s)$ and $\alpha_{\pm}$
are given in equation (\ref{alph}). The initial equal-time responses $R(s,s)$ and 
$Q(s,s)$ depend on the functions $C^{gf}_m(s)$, but for the chosen KDH dynamics,
closed equations for them are unknown. In any case, the responses $R(t,s)$ and 
$Q(t,s)$ should decay, at large times, as $\exp(-(t-s)/\tau_{-})$.

Aspects of non-equilibrium relaxation can be explicitly studied for responses with
respect to the initial state, since closed expressions can be given for $R(t,0)$
and $Q(t,0)$. We study three examples:
\begin{enumerate}
\item Consider fully ordered initial states $\cdots\uparrow\uparrow\uparrow\uparrow\cdots$ and 
$\cdots\downarrow\downarrow\downarrow\downarrow\cdots$ at the critical point $\delta=1$. 
Then both $R(0,0)=Q(0,0)=0$ and these global responses $R(t,0)=Q(t,0)=0$ 
of the system will not react to a small perturbing external field. 
Hence there is no perceptible 
{\em equilibrium} relaxation at criticality via global observables for $1D$ 
KDH dynamics. 
\item In order to study {\em non}-equilibrium relaxation, consider a fully disordered
initial state and quench at time $t=0$ the control parameter to the value $\delta$. Then $R(0,0)=1$ and $Q(0,0)=\delta$. For the limit $\delta\to 1$, we find $R(t,0)=1$.
This indeed describes a {\em non}-equilibrium relaxation, as may be seen by
considering an initial state at thermal equilibrium with temperature $T_{\rm ini}>0$. Standard techniques \cite{Thompson72} may be used to calculate $R(0,0)$ and $Q(0,0)$ and we find that for $\delta=1$
\BEQ \label{gl:Rtempini}
R(t,0) = \bigl( 1 - \tanh \eta\bigr)^3 
+ \tanh\eta \bigl( 1-\tanh\eta\bigr)^2\, e^{-2t}
\EEQ
where we have set $\eta:=J/T_{\rm ini}$. The stationary value of the response function is distinct from the equilibrium value, which would only be reached if the
limit $\eta\to\infty$ were taken.  
\item In order to appreciate better the role of the initial state in non-equilibrium relaxation, consider the initial ensemble made from the states
$\cdots\uparrow\uparrow\downarrow\uparrow\uparrow\downarrow\uparrow\uparrow\downarrow\cdots$ and 
$\cdots\downarrow\downarrow\uparrow\downarrow\downarrow\uparrow\downarrow\downarrow\uparrow\cdots$ (with equal probability) 
and couple the system at the initial time to a heat bath such that
the control parameter has the value $\delta$. Then $R(0,0)=1+\delta/3$ and $Q(0,0)=-1+5\delta$. We find in the limit $\delta\to 1$ that 
$R(t,0)= (4/3)\exp(-2t)$ relaxes exponentially fast towards its equilibrium value. 
\end{enumerate}
For an interpretation in terms of dynamical scaling, we recall that the global
responses calculated here are actually the Fourier transforms, viz. 
$R(t,s) = \left. \int_{\mathbb{R}}\!\D r\: e^{-\II q r} R(t,s;r,0) \right|_{q=0}$ 
of the space-time responses defined in the introduction, at vanishing momentum
$q=0$. Hence from (\ref{gl:scalinf}) we expect
$R(t,0) = t^{(1-\lambda_R)/z} \int_{\mathbb{R}} \!\D u\: \Phi_R(u)$. 
For $t$ sufficiently large, this may be compared in
particular with equation (\ref{gl:Rtempini}), and 
this scaling interpretation suggests that
\begin{equation}
\lambda_R=1.
\end{equation}
Provided that $\alpha_+ R(0,0)-\delta Q(0,0)\ne 0$, this result is generic.
The independence of the initial state confirms the expected universality of the
autoresponse exponent $\lambda_R$.

\subsubsection{Correlation functions:}
The equations of motion for the global correlation functions $C^{gf}_{r}(t)$ and $C^{gf}(t)$ are explicitly obtained in the appendix, see eqs.~(\ref{ssC-g},\ref{sqC-g},\ref{qqC-g}). While in general a closed system of
equations of motion cannot be found, it turns out if the initial conditions $C^{gf}(0)$ are of order ${\rm O}(1)$, then in the limit $L\to \infty$ the
{\em global} correlators $C^{gf}(t)$ satisfy the following closed set of
linear equations 
\begin{eqnarray}
\label{eom-C}
\frac{\D}{\D t} 
\left( \begin{array}{c} 
 C^{\sigma \sigma}(t) \\C^{q\sigma}(t) \\ C^{qq}(t)
\end{array} \right)
=
\left( \begin{array}{ccc} 
4\delta -2 &  - 2\delta & 0 \\
3\delta & 2\delta -4 &  - \delta \\
0& 6\delta & - 6
\end{array} \right)
\left( \begin{array}{c} 
 C^{\sigma \sigma}(t) \\C^{q\sigma}(t) \\ C^{qq}(t)
\end{array} \right).
\end{eqnarray}
Upon resolving, we find the global spin-spin fluctuations to be
\begin{eqnarray}
\label{sol-C-ss}
C^{\sigma \sigma}(t) = 
B_{-}e^{-2\lambda_{-} t} + B_{0}e^{-(\lambda_{-}+\lambda_{+}) t} + B_{+} e^{-2\lambda_{+} t}
\end{eqnarray}
where $\lambda_{\pm}$ are given in (\ref{eigen-lam}) and 
\begin{eqnarray}
\label{B-def}
B_{\mp} :=  \frac{1}{4\Delta^2}
\left( \alpha^2_{\pm} C^{\sigma \sigma}(0) -2\delta \alpha_{\pm} C^{q\sigma}(0)
+ \delta^2 C^{qq}(0) \right), \\
B_{0} :=  -\frac{1}{2\Delta^2}
\left( \alpha_{+}\alpha_{-} C^{\sigma \sigma}(0) -2\delta (1+\delta) C^{q\sigma}(0)
- \delta^2 C^{qq}(0) \right).
\end{eqnarray}
The explicit expressions for the other two correlation functions are
\begin{eqnarray}
\label{sol-C-qs}
C^{q\sigma}(t) =  \frac{1}{\delta} \left(
\alpha_{-}B_{-}e^{-2\lambda_{-} t} + (1+\delta)B_{0}e^{-(\lambda_{-}+\lambda_{+}) t} + 
\alpha_{+}B_{+} e^{-2\lambda_{+} t}
\right),
\\
\label{sol-C-qq}
C^{qq}(t) =  3\left(
\frac{\alpha_{-}}{\alpha_{+}} B_{-}e^{-2\lambda_{-} t} 
+ B_{0}e^{-(\lambda_{-}+\lambda_{+}) t} + 
\frac{\alpha_{+}}{\alpha_{-}} B_{+} e^{-2\lambda_{+} t}
\right).
\end{eqnarray}
The global correlations at large times in general relax as $\exp(-2t/\tau_{-})$. Since the leading correction, coming form the non-global terms, to these correlations will be of order ${\rm O}(t/L)$, the solution is valid for any time $t < \tau_{-}$ in the large-$L$ limit only for $(B_{0}, B_{\pm})$ of order ${\rm O}(1)$.

The two-time correlation functions of $\sigma_n$ and $q_n$ variables for 
$t \ge s^+$ is denoted by
\begin{equation}
\label{2-t-def0}
C^{gf}_{n,m}(t,s)_{+} := \sum_{\sigma,\widetilde{\sigma}} g_n \widetilde{f}_m P(\sigma,t| \widetilde{\sigma},s),
\end{equation}
where $f_n$ and $g_n$ are either $\sigma_n$ or $q_n$. These quantities have similar equations of motion as those of one-point functions of $\sigma_n$ and $q_n$. For any time $t$ the correlation functions are defined as
\begin{eqnarray}
\label{2-t-def}
{C}^{gf}_{n,m}(t,s) 
=
\left\{ \begin{array}{c} 
C^{gf}_{n,m}(t,s)_{+}   ~~ \mbox{for} ~~ t \ge s^+\\
C^{fg}_{m,n}(s,t)_{+}   ~~ \mbox{for}  ~~t \le s^-\\
\langle g_n f_m \rangle_s   ~~~\mbox{for}  ~~t = s
\end{array} \right.
\end{eqnarray}
The two-time correlation function of corresponding global quantities are defined by 
\begin{equation}
\label{2-t-def-glob}
C^{gf}(t,s) :=  \frac{1}{L^2}\sum_{m,n} C^{gf}_{n,m}(t,s), 
\end{equation}
and the equations of motion for $t \ge s^+$ are given by the following familiar set,
\begin{eqnarray}
\label{eom-C-2t}
\frac{\partial}{\partial t} 
\left( \begin{array}{c} 
C^{\sigma f}(t,s)_{+}  \\ C^{q f}(t,s)_{+}  
\end{array} \right)
=
\left( \begin{array}{cc} 
2\delta -1 &  - \delta  \\
3\delta & -3  
\end{array} \right)
\left( \begin{array}{c} 
C^{\sigma f}(t,s)_{+}  \\ C^{q f}(t,s)_{+}  
\end{array} \right) .
\end{eqnarray}
The general solution for these correlation functions in the regime $t \ge s^+$ is given by
\begin{eqnarray}
\label{sol-2t-1}
C^{\sigma f}(t,s) &=& 
 A^{\sigma f}(s) e^{-\lambda_{-}(t+s)}  + E^{\sigma f}(s) e^{-(\lambda_{-}t
+\lambda_{+}s)} \nonumber \\
& & + F^{\sigma f}(s) e^{-(\lambda_{+}t+\lambda_{-}s)} 
+ B^{\sigma f}(s) e^{-\lambda_{+}(t+s)} ,
\\
\label{sol-2t-2}
C^{q f}(t,s) &=&  \frac{1}{\delta}\alpha_{-}
 \left( A^{\sigma f}(s) e^{-\lambda_{-}(t+s)}  + E^{\sigma  f}(s) e^{-(\lambda_{-}t
+\lambda_{+}s)} \right)
\nonumber \\
& & + \frac{1}{\delta} \alpha_{+}\left(
F^{\sigma  f}(s) e^{-(\lambda_{+}t+\lambda_{-}s)} 
+ B^{\sigma  f}(s) e^{-\lambda_{+}(t+s)} \right) ,
\end{eqnarray}
where  $A^{\sigma  f}(s)$, $B^{\sigma  f}(s)$, $E^{\sigma  f}(s)$, and  $F^{\sigma  f}(s)$ are arbitrary functions of $s$. These functions are fixed if we require that at $t=s$ the two-time correlation functions become the equal-time correlation functions as given in equations (\ref{sol-C-ss}), (\ref{sol-C-qs}), and (\ref{sol-C-qq}). In fact, all the arbitrary functions turn out to be independent of $s$, and are explicitly 
given by
\begin{eqnarray}
\label{arb-fun}
A^{\sigma \sigma} = B_{-} ~,~
B^{\sigma \sigma} = B_{+} ~,~
A^{\sigma q} = \frac{1}{\delta}\alpha_{-}B_{-} ~,~
B^{\sigma q} = \frac{1}{\delta} \alpha_{+}B_{+} ~,~ \nonumber \\
E^{\sigma \sigma} = F^{\sigma \sigma}  = \frac{1}{2} B_{0} ~,~
E^{\sigma q} =\frac{1}{2\delta}\alpha_{+}B_{0} ~,~
F^{\sigma q}  = \frac{1}{2\delta}\alpha_{-}B_{0} .
\end{eqnarray}
For these values, the expression for $C^{\sigma q}(t,s)_{+}$ in equation~(\ref{sol-2t-1}) coincides with that for
$C^{q \sigma}(s,t)_{+}$ in equation~(\ref{sol-2t-2}), and hence the piecewise solution in $t \ge s$ extends to the $t \le s^-$ region, too. 

The equations of motion (\ref{eom-C-2t}) for the two-time correlations are exact for periodic boundary conditions. On the other hand, the system of equations (\ref{eom-C}) that describes the equal-time correlations only holds true in the infinite-size limit $L\to\infty$ only when at least one the global correlators is much larger than of order ${\rm O}(1/L)$. Therefore, the global correlations with the
initial state are given by 
\begin{eqnarray}
\label{sfC-t0}
C^{\sigma f}(t,0)= \frac{1}{2\Delta} \left( A_{-}^{\sigma f} e^{-\lambda_{-}t} - A_{+}^{\sigma f} e^{-\lambda_{+}t} \right)
\\
C^{q f}(t,0)= \frac{1}{2\Delta \delta} \left(\alpha_{-} A_{-}^{\sigma  f} e^{-\lambda_{-}t} - \alpha_{+}A_{+}^{\sigma  f} e^{-\lambda_{+}t} \right)
\end{eqnarray}
where $A_{\mp}^{\sigma f} = \alpha_{\pm} C^{\sigma f}(0) - \delta C^{q f}(0) $. 

We now discuss the behaviour of the global two-time correlators for the same
initial states as discussed above for the response function.
\begin{enumerate}
\item For the fully ordered initial states one obtains equilibrium relaxation at the static critical point $\delta=1$. We have $B_-=B_0=1$ and $B_+=0$ and find
\BEQ
C^{\sigma\sigma}(t,s) = 1 + \frac{1}{2} \left( e^{-2s}+e^{-2t}\right)
\EEQ
such that one has an exponentially fast relaxation towards the global equilibrium
correlator $C_{\rm eq}^{\sigma\sigma}=1$. 
\item For a fully disordered state the equations of motion do not close and
(\ref{eom-C}) are not valid. However, the global correlator with the initial 
state may be read off from (\ref{sfC-t0}) and we find $C^{\sigma\sigma}(t,0)=1/L$. 
Similar results may be found for any initial temperature $T_{\rm ini}>0$. 
\item For the partially ordered initial states $\cdots\uparrow\uparrow\downarrow\uparrow\uparrow\downarrow\uparrow\uparrow\downarrow\cdots$ and 
$\cdots\downarrow\downarrow\uparrow\downarrow\downarrow\uparrow\downarrow\downarrow\uparrow\cdots$ we find at the critical point $\delta=1$ that
$B_-= 1$, $B_0= -1/3$ and $B_+= 4/9$. Hence
\BEQ \label{gl:C2ts}
C^{\sigma\sigma}(t,s) = 1 - \frac{1}{6}\left( e^{-2s}+e^{-2t}\right)
+ \frac{4}{9} e^{-2(t+s)}
\EEQ
which, again, relaxes exponentially towards equilibrium. For the global spin-spin correlator with the initial state we find from (\ref{sfC-t0})
\BEQ \label{gl:C2t0}
C^{\sigma\sigma}(t,0) = \frac{1}{3} - \frac{2}{9} e^{-2t}
\EEQ
which is distinct from the formal $s\to0$ limit of the two-time result (\ref{gl:C2ts}). 
\end{enumerate}
For an interpretation in terms of the scaling form (\ref{gl:scal}), recall that 
$C^{\sigma\sigma}(t,s)$ actually is the Fourier transform
\BEQ \label{gl:C2skal}
\hspace{-1.2truecm}C^{\sigma\sigma}(t,s)=\left.\int_{\mathbb{R}}\!\D r\: e^{-\II q r} C(t,s;r,0)\right|_{q=0} = s^{-b+1/z} \left(\frac{t}{s}-1\right)^{1/z} \int_{\mathbb{R}} \!\D u\, F_C\left( u, \frac{t}{s}\right)
\EEQ
and similarly, a correlation with respect to the initial state is interpreted via
(\ref{gl:scalinf}) as
\BEQ
\hspace{-1.2truecm}C^{\sigma\sigma}(t,0) = \left.\int_{\mathbb{R}}\!\D r\: e^{-\II q r} C(t,0;r,0)\right|_{q=0} = t^{(1-\lambda_C)/z} \int_{\mathbb{R}}\!\D u\: \Phi_C(u)
\EEQ
Comparing with the explicit result (\ref{gl:C2t0}) for large times, we read off
\BEQ
\lambda_C =1.
\EEQ
If we could formally compare eqs.~(\ref{gl:C2ts},\ref{gl:C2skal}), we would find
$b=1/z=1/4$, but since (\ref{gl:C2ts}) apparently describes a relaxation towards
{\em equilibrium}, it is not clear whether the non-equilibrium 
scaling form (\ref{gl:C2skal}) is applicable in the present context. 

\section{Conclusions}
\label{conclude}

The exact study of one-dimensional kinetic Ising models  may provide useful
insight into the slow relaxation behaviour of many-body systems with strongly
interacting degrees of freedom.\footnote{In order to design new slowly relaxing nanosystems, with a view of possible applications to information storage, the slow relaxation dynamics in real systems such as the single-chain magnet [Mn$_2$(saltmen)$_2$Ni(pao)$_2$(py)$_2$](ClO$_4$)$_2$ has been explicitly 
compared to $1D$ Glauber dynamics \cite{Coulon04}.}  When considering the local spin-flip dynamics
given by the rates (\ref{Ws}), the KDH line $\gamma=2\delta$ is an interesting
variant of the celebrated Glauber dynamics $\delta=0$~\cite{Glauber63}. 
We have generalised the known exact results for the global magnetisation and the global three-spin magnetisation~\cite{Deker79,Kimball79} and have shown how certain global two-time correlators and certain global response functions (susceptibilities) may be found, when suitable initial conditions are chosen.  

Although the KDH dynamics does satisfy detailed balance, the dynamical behaviour
close to the critical point at zero temperature is different from the one found
for Glauber dynamics, since at $T=0$ the number of stationary states grows
exponentially with the number $L$ of lattice sites. This property might be
seen as an analogy with the many (meta-)stable states in glassy or kinetically
constrained systems, such as the Frederikson-Andersen model, see e.g. \cite{Frederikson84,Mayer06}, or in frustrated magnets, see e.g.  \cite{Wannier50,Han08,Walter08}. In particular, analysis of the critical slowing-down near criticality gives the KDH dynamical exponent~\cite{Deker79}
\BEQ
z=4
\EEQ
in contrast to the result $z=2$ of Glauber dynamics \cite{Glauber63}. 
We studied the non-equilibrium relaxation of KDH dynamics through the global
two-time correlators and responses. Although our results appear to be compatible 
with the usually admitted scaling behaviour (\ref{gl:scalinf}) and we have
in this way identified some non-equilibrium exponents,
\BEQ
\lambda_C = \lambda_R =1,
\EEQ
further tests of non-equilibrium dynamical scaling in the KDH-Ising model 
are desirable.
The ageing exponent $a$ could not be determined. Before accepting the formal conjecture $b=1/z=1/4$, the applicability of the scaling form (\ref{gl:scal}) needs
to be checked, which cannot be done from the present analytical results alone. 
These values of exponents should be broadly independent of the precise form of the initial states, in agreement with the expected universality. We observe that
the values of $\lambda_C=\lambda_R$ agree with what is found for Glauber dynamics, while the conjectured value of $b$ is different from the Glauber dynamics result $b=0$, see \cite{Godreche00a,Lippiello00,Henkel04}. 

Since there is a duality mapping onto a diffusion-annihilation process, it would
be interesting to see how the results obtained here might bear on that system. 
In this context, we remark that in a similar way as analysed in this paper, 
one can derive closed systems of equations of motion for {\em staggered} quantities like $M_s$ and $T_s$, but along a different line $\gamma=-2\delta$. This line
can also be obtained from the line $\gamma=2\delta$ via a gauge transformation $\sigma_n \rightarrow (-1)^n \sigma_n$  and $\gamma \rightarrow -\gamma $, which is usually made use to relate ferromagnetic and anti-ferromagnetic Ising models. Graphically, this can be illustrated in figure~\ref{Fig2}, where the `anti-ferromagnetic' case $(-\gamma,\delta)$ can be obtained by a reflection about the $\gamma=0$ line. Quantities such as $M_s(t)$ have an immediate interpretation as the total particle number in the dual diffusion-annihilation process.

\appsektion{Equations of motion for global correlation functions}
We derive the equations of motion for the correlation functions $C^{gf}_r(t)$ and $C^{gf}(t)$ along KDH line $\gamma = 2 \delta$. Here, we need not assume that $C^{gf}_r(t)= C^{fg}_r(t)$, although for $r=1,2$ it can be easily shown that $C^{\sigma q}_1(t)= C^{q \sigma}_1(t)$ and $C^{\sigma q}_2(t)= C^{q \sigma}_2(t)$. 

The equation of motion for $C^{\sigma \sigma}_r(t)$, for $r \ne 0$, is obtained from equation (\ref{s-s}) and reads
\begin{eqnarray}
\label{ssC-r}
\hspace{-0.5truecm}\frac{\D}{\D t}  C^{\sigma \sigma}_r(t) = -2 C^{\sigma \sigma}_r(t) 
+ 2 \delta \left( C^{\sigma \sigma}_{r-1}(t) + C^{\sigma \sigma}_{r+1}(t) \right)
-\delta  \left( C^{\sigma q}_r(t) + C^{q \sigma}_r(t)\right).
\end{eqnarray}
For $r=0$, we have $C^{\sigma \sigma}_0(t)=1$.
The dynamic equation for $C^{\sigma q}_r(t)$, for $r \ne 0,\pm 1$, obtained from equation (\ref{q-s}), is given by
\begin{eqnarray}
\label{sqC-r}
\frac{\D}{\D t}  C^{\sigma q}_r(t) = &-&4 C^{\sigma q}_r(t) 
+ 2 \delta \left( C^{\sigma \sigma}_{r-1}(t) + C^{\sigma \sigma}_{r+1}(t) \right)
-\delta  \left( C^{\sigma \sigma}_r(t) + C^{q q}_r(t)\right)
\nonumber \\
&+& \delta \left( C^{\sigma q}_{r-1}(t) + C^{\sigma q}_{r+1}(t) \right) + \delta C^{\sigma A}_r(t),
\end{eqnarray}
where $C^{\sigma A}_r(t) := \sum_n \langle \sigma_n A_{n+r}\rangle/L$. This is supplemented by 
\BEQ
\label{sqC-0}
C^{\sigma q}_0(t)= C^{\sigma \sigma}_2(t) \;\; , \;\; 
\label{sqC-1}
C^{\sigma q}_{\pm 1}(t)= C^{\sigma \sigma}_1(t) .
\EEQ
The equation of motion for $C^{q q}_r(t)$, for $r \ne 0, \pm1, \pm2$, is obtained from equation (\ref{q-q}) and is given by
\begin{eqnarray}
\label{qqC-r}
\frac{\D}{\D t}  C^{q q}_r(t) = &-&6 C^{q q}_r(t) 
+ 2 \delta \bigl( C^{q \sigma}_{r-1}(t) + C^{\sigma q}_{r-1}(t) \bigr)
+ 2 \delta \bigl( C^{q \sigma}_{r+1}(t) + C^{\sigma q}_{r+1}(t) \bigr)
\nonumber \\
&-&\delta  \left( C^{q \sigma}_r(t) + C^{\sigma q}_r(t) \right)
+ \delta \left( C^{q A}_r(t) + C^{A q}_r(t) \right), 
\end{eqnarray}
while that for  $C^{q q}_2(t)$ is obtained from equation (\ref{q-q-2}) and reads
\begin{eqnarray}
\label{qqC-2}
\frac{\D}{\D t}  C^{q q}_2(t) = &-&4 C^{q q}_2(t) 
+ 4 \delta C^{\sigma \sigma}_1(t) + 2 \delta C^{\sigma \sigma}_4(t) 
- 2 \delta  C^{\sigma q}_2(t) 
\nonumber \\
&+& \delta \left( C^{q A}_2(t) + C^{A q}_2(t) \right).
\end{eqnarray}
For $r = 0, 1$, we have 
\BEQ
\label{qqC-0}
C^{q q}_0(t)= 1 \;\; , \;\; 
\label{qqC-1}
C^{q q}_{1}(t)= C^{\sigma \sigma}_3(t) .
\EEQ
The equations of motion for $\{ C^{\sigma \sigma}_r, C^{q \sigma}_r,  C^{q q}_r\}$ are not closed as they also involve $C^{A q}_r$ terms. Hence explicit solutions
can only be found via approximate methods, such as mean-field theories or truncations.

However, one may obtain the equations of motion for the {\em global} correlation functions $C^{gf}(t)$. In the expansion
\begin{equation}
\label{expan-1}
\frac{\D}{\D t}  C^{\sigma \sigma}(t) = \frac{1}{L^2}\sum_{n} \sum_{m \ne n} \frac{\D}{\D t} \langle \sigma_n \sigma_m \rangle ,
\end{equation}
upon substituting equation (\ref{s-s}) and rewriting in terms of correlation functions, we obtain
\begin{eqnarray}
\label{ssC-g}
\hspace{-0.5truecm}\frac{\D}{\D t}  C^{\sigma \sigma}(t) = (4\delta -2)  C^{\sigma \sigma}(t) - 2\delta C^{q\sigma}(t) 
+ \frac{2}{L} \Big( 1 \!-\! 2\delta C^{\sigma \sigma}_{1}(t) \!+ \! \delta  C^{ \sigma \sigma}_{2}(t) \Big).
\end{eqnarray}
Similarly, in the expansion
\begin{equation}
\label{expan-2}
\hspace{-0.5truecm}\frac{\D}{\D t}  C^{\sigma q}(t) = \frac{1}{L^2}\sum_{n} \sum_{m \ne n, n\pm1} \frac{\D}{\D t} \langle q_m\sigma_n \rangle + \frac{1}{L}  \left( \frac{\D}{\D t} C^{\sigma q}_0(t)+ 2 \frac{\D}{\D t} C^{\sigma q}_1(t)\right)
\end{equation}
upon using equations (\ref{q-s}), (\ref{sqC-0}), 
and (\ref{ssC-r}), we find
\begin{eqnarray}
\label{sqC-g}
\lefteqn{ \hspace{-0.5truecm}
\frac{\D}{\D t}  C^{\sigma q}(t) = 3\delta   C^{\sigma \sigma}(t) + (2\delta -4) C^{\sigma q}(t) -\delta C^{q q}(t) 
}
\nonumber \\
& & + \frac{2}{L} \Big(\delta +  2(1\!-\!\delta) C^{\sigma \sigma}_{1}(t) +
(1\!-\!2\delta)  C^{ \sigma \sigma}_{2}(t) 
+  2\delta  C^{ \sigma \sigma}_{3}(t) -  2\delta  C^{ \sigma q}_{2}(t)\Big).
\end{eqnarray}
Finally, from the expansion
\begin{equation}
\label{expan-3}
\hspace{-0.5truecm}\frac{\D}{\D t}  C^{q q}(t) = \frac{1}{L^2}\sum_{n} \sum_{m \ne n, n\pm1, n\pm2} \frac{\D}{\D t} \langle q_m q_n \rangle + \frac{2}{L} \left( \frac{\D}{\D t} C^{q q}_1(t)+ \frac{\D}{\D t} C^{q q}_2(t)\right)
\end{equation}
and upon using equations (\ref{q-q}), (\ref{qqC-1}), (\ref{qqC-2}), and (\ref{ssC-r}), we obtain
\begin{eqnarray}
\label{qqC-g}
\lefteqn{ \hspace{-0.5truecm} 
\frac{\D}{\D t}  C^{q q}(t) = 6\delta  C^{\sigma q}(t) - 6 C^{q q}(t)
}
\\
& & + \frac{2}{L} \Big(3 -2\delta C^{\sigma \sigma}_{1}(t) -\delta  C^{ \sigma \sigma}_{2}(t) 
+  4 C^{ \sigma \sigma}_{3}(t) 
+ 2\delta C^{ \sigma \sigma}_{4}(t) 
\nonumber \\
& & 
-  4 \delta  C^{ \sigma q}_{2}(t) - 2 \delta \bigl( C^{ \sigma q}_{3}(t)+ C^{q \sigma}_{3}(t)\bigr)
+ 2   C^{q q}_{2}(t)
\Big).
\nonumber
\end{eqnarray}
The equations of motion for $\{ C^{\sigma \sigma}, C^{q \sigma},  C^{q q} \}$ also involve $\{ C^{\sigma \sigma}_{1}, C^{\sigma \sigma}_{2},C^{\sigma \sigma}_{3},C^{\sigma \sigma}_{4}, C^{ \sigma q}_{2}, C^{ \sigma q}_{3}, C^{q q}_{2} \}$. Although these equations are not closed in general, they sometimes do close 
in the thermodynamic limit, at least for certain initial conditions. 

For example, in the case of a fully disordered initial state,  
$C^{gf}_r$ are of order ${\rm O}(1/\sqrt{L})$, while  $C^{gf}$ are of order ${\rm O}(1/L)$. Now the solutions $C^{gf}_r(t)$ will be of the form $\mathfrak{f}^{gf}_r(t)+ \mathfrak{g}^{gf}_r(t)/\sqrt{L}$. 
This set of functions $\{ \mathfrak{f}^{gf}_r(t)\}$ is needed to compute $C^{gf}(t)$ to order ${\rm O}(1/L)$, but for which no closed system of equations is known. On the other hand, if we choose an initial state such that 
$C^{gf}(0) = {\rm O}(1)$, the limit $L \to \infty$ in equations (\ref{ssC-g}, \ref{sqC-g}, \ref{qqC-g}) can be taken and the resulting closed system (\ref{eom-C}) may be solved explicitly, as discussed in the main text. 

~\\
\noindent
{\bf Acknowledgements:}
This work was supported by the Franco-Korean binational exchange programme PHC Star, N$^{o}$ 16546TD. 
MH thanks KIAS for warm hospitality, where this work was done. SD and HP acknowledge 
the generous support by the people of South-Korea, including that by the Korea 
Foundation for International Cooperation of Science and Technology (KICOS) through a 
grant provided by the Korean Ministry of Science and Technology (MOST) 
with No. 2007-00369.

\end{document}